# Building Effective Virtual Teams: How to Overcome the Problems of Trust and Identity in Virtual Teams

Chris Kimble

**Abstract**
This article explores some of the challenges faced when managing virtual teams, in particular the role played by trust and identity in virtual teams. It outlines why teams and virtual teams have become a valuable part of the modern organization and presents ten short case studies that illustrate the range of activities in which virtual teams can be found. Following this, the article examines some of the common problems encountered in virtual team working. It discusses two broad classes of solutions. The first are solutions that are essentially technical in nature (i.e., where changes to or improvements in technology would help to solve or ameliorate the problem); the second are more organizationally based (i.e., where the root of the problem is in people and how they are managed). The article concludes that both the technical and the organizational solutions need to be considered in parallel if an attempt to build an effective virtual team is to be successful.

## 1   Introduction

Workgroups of various sorts are the fundamental building blocks of the organization. In the traditional organization, these groups would be functional departments, like sales, engineering, or manufacturing. However, in recent years, many companies have begun to move toward a style of working that is explicitly cross-functional and built upon flatter organizational structures. Instead of the traditional functional areas and hierarchies, these companies are moving toward team-based structures, where groups of people take responsibility for a particular organizational deliverable. In some cases, these may be whole products or services; in other cases, they may be sub products or some other element of the organization's value chain.

In an era of increasing globalization, international trade, and fast communications networks, these team-based structures may consist of members that are located in different buildings, different cities, or even different continents. Teams that operate in this way are sometimes called "virtual teams." This article will discuss some of the challenges that organizations face when managing virtual teams and, in particular, will examine the role played by trust and identity. It will explore how organizations respond to these challenges under two broad headings: those that are principally "technical" problems in the sense that technology can help to solve or reduce the effect the problem has and those that are "organizational," where the solution to the problem lies mainly with people and how they are managed, before going on to discuss how these challenges should be dealt with in practice.

## 2   Teams

Teams are organizational units that share a common goal and whose members have a feeling of mutual responsibility for the results that the team produces. Thus, to some extent, teams are self-managing, as to work in this way requires a degree of mutual commitment if the team is to achieve its goal. Because teams are not fixed, they can be more easily formed and disbanded than traditional hierarchical structures; thus, teams and team working offer a flexible, dynamic, and efficient way of organizing working life.

Teams also allow members to share information that would previously have never crossed the walls of the traditional functional silos based on departments.

In essence, teams provide vehicles for learning from others. To learn, people need time to reflect and a safe environment. By working together over a period of time, team members are able to develop a sense of trust and shared identity that increases their ability to share and learn from each other. If a team can develop a sense of trust and mutual respect, then people feel able to share their thinking, the reasons behind their conclusions, and even the doubts that they have about their conclusions. Together they can build on each other's ideas, create new ideas, and develop new insights.

Team-working strategies have been used in a variety of industries, from mobile phones to banking and financial institutions. Some of the earliest examples of this can be found in manufacturing, where teams of multi-skilled operators worked together to build a complete subassembly (Gyllenhammar, 1977). However, the same phenomenon can be found in product development, where people from different areas of the company work together to design new products, and in the service sector, where people with different backgrounds work together to provide a range of different services for a client.

## 3  Virtual Teams

Although examples of the move toward team-based organizational structures can be found dating back to the 1970s, the nature of modern teamwork has changed significantly. Thanks in part to the power of information technology and fast and reliable communications networks, organizations have become more geographically distributed, and firms have increasingly begun to work in partnerships that span different industries. Where once teams were about relationships inside an organization, now they frequently include those that were previously considered to be outside, such as customers, suppliers, and other organizations with whom they collaborate. For example, Kimble, Grenier, and Goglio-Primard (2010) describe how two firms who were potentially competitors were able to cooperate with each other and their clients to build a customer relationship management system that neither would have been able to build on its own.

The term virtual team can be applied to a number of different types of groups. Team membership may be relatively stable (e.g., in an established sales team) or change on a regular basis (e.g., in project teams). Members may be drawn from the same organization or from several different organizations (e.g., when projects involve consultants or external assessors). Team members may work in close proximity (e.g., in the same building) or geographically distantly (e.g., in different countries) and, similarly, team members may work at the same or at different times (e.g., depending on whether the team members are in the same time zone).

However, despite the widespread use of mobile telephones, computers, and the Internet, truly virtual teams, in the sense that they only ever "meet" together through technology, are relatively rare. A far more common scenario is one where team members continue to engage in face-to-face contact of one form or another throughout the life of the team. Most "virtual" teams operate in multiple modes: sometimes face to face, sometimes via electronic communication, sometimes interacting with each other directly, and sometimes working as individuals. Managing such teams means managing a whole spectrum of different communication strategies and techniques, as well as managing the human and social processes that nourish and support the esprit de corps that makes a team a team.

However, before we go on to look at the problems of managing virtual teams, we will first look at some case studies that illustrate the range of virtual teams that can be found in different sectors (Kimble, Li, & Barlow, 2000). The information about these virtual teams was collected through face-to-face interviews, combined with other forms of correspondence such as e-mails, faxes, company reports, and phone calls.

Exhibit 1. The Case Studies

|  | **Main activity** | **Intra or inter-organizational team** | **Synchronous or asynchronous communication** | **Geographical spread** | **Number of geographical locations** |
|---|---|---|---|---|---|
| **Case 1** | Software support | Inter-organizational | Synchronous | National / International | 3-5 |
| **Case 2** | Software development | Both | Both | National | 2-5 |
| **Case 3** | Software development | Intra-organizational | Both | Regional | 3-8 |
| **Case 4** | Legal services | Intra-organizational | Mostly synchronous | Regional | 3 |
| **Case 5** | Secretarial services | Intra & Inter-organizational | Synchronous | Local | 2-4 |
| **Case 6** | Research and consultancy | Inter-organizational | Both | Regional | 2-5 |
| **Case 7** | Home based market research | Intra-organizational | Both | Local | 2-5 |
| **Case 8** | Hospital and medical services | Inter-organizational | Synchronous | Regional | 2 |
| **Case 9** | Hospital and medical services | Inter-organizational | Synchronous | Regional | 2-3 |
| **Case 10** | Enquiries for home based workers | Intra-organizational | Both | Regional | 2 + |

## 4   Virtual Teams in Practice: Some Illustrative Case Studies

The first example (see case 1 in Exhibit 1) is a virtual team that involves a CASE (Computer-Aided Software Engineering) tool supplier and its main customer in the United Kingdom. The supplier develops a range of software applications for customers in aerospace and defense, telecommunications, electronics, systems software, and manufacturing. As part of the services it provides, it offers constant, high-quality technical support to its customers. In the past, this was provided by the supplier sending experts directly to customers' sites, but now a virtual team-based solution has enabled the company to find a more effective way of supporting its customers as well as provide them with greater responsiveness and flexibility.

The task of supporting these systems is a complex undertaking, potentially involving interactions at many different technical and managerial levels, and requiring high levels of interactivity among geographically dispersed individuals. In order to provide the level of support that is required, high specification workstations, which have the ability to support a wide range of applications simultaneously, need to be networked together because problem solving often requires simultaneous step-by-step operations on CASE tools at both ends of a link. A multimedia communications link, with audio-visual capability, supports direct interaction between users and experts. Taking this approach, the geographical flexibility of the experts and the responsiveness of the service has improved significantly, especially in the case of mission-critical situations such as a systems breakdown.

The system now covers three sites: the head office in southeast England, a remote office in the north of England, and a large customer site in the West Midlands. The services offered by the company include tele-support for tool use (such as installation, problem identification, and decisions about tool-use and tool-method issues), tele-guidance for systems engineering methods (such as queries, method teaching, and modeling alternatives), and tele-reviews of analysis and design models (via synchronized model browsing).

Similar applications were identified in two other companies: one in a team of software developers in Northern Ireland who develop software remotely for a client in London (case 2 in Exhibit 1) and the other in a team of software engineers in Scotland (case 3 in Exhibit 1) who work from home on a variety of software-development projects. In all three cases, both the audio-visual and data communications facilities of the systems were essential if the teams were to work together effectively.

A different approach to virtual teams was identified in a large law firm in Germany with multiple offices (case 4 in Exhibit 1). The provision of the full range of professional legal services in remote locations, with only small branch offices and a limited number of clients, is expensive. The result is a lower level of service in rural areas. This often has a self-reinforcing effect, since poorer-quality services discourage firms that have more specialized legal needs from locating in these areas. However, this situation has been alleviated to some extent by the use of dedicated telecommunications networks that support the provision of legal services in branch offices through providing direct communication links to larger offices in more central locations.

In this case, the solution was developed between a main office and two branch offices in northern Germany. The intention was not only to provide enhanced services to remote locations but also to reverse the trend for experienced legal experts to migrate to the cities by offering them the possibility of providing their services from more remote rural locations. The idea behind the innovation was that a particular legal expert would not have to be based in the main office but could provide services from a branch office. This application required reasonable quality videophones that could support the sort of direct exchanges one might find in any legal consultancy. In addition, the system also needed to support the simultaneous viewing of images or documents, and the transmission of large volumes of case file data. Despite various difficulties, the system has significantly improved the geographical flexibility of the firm's legal experts and the quality and responsiveness of the firm's service to customers.

Similar virtual teams were identified in other sectors. In France, a business services company (case 5 in Exhibit 1) set up a system to support communication among its central office in Paris, three satellite offices in the suburbs, and several regular clients. The system enabled direct communication, and the simultaneous viewing and editing of word-processed documents, through an audio-visual data link with clients. This allowed complex editing and formatting issues on documents to be quickly resolved. Direct communications between the central and satellite offices of the company were also used for the allocation and scheduling of work. This system helped to improve some existing problems, such as errors and delays in correcting typed documents, which were major elements constraining the further expansion of the company.

In southern Italy, a system was developed to link together several academic and research institutions to provide a full range of research, training, and consultancy services for industry (case 6 in Exhibit 1). The system they use supports the transmission of large volumes of multimedia information (including audio, images, graphics, text, and data) in real time. Another example can be found in a market research firm (case 7 in Exhibit 1). Here a new system was developed to support collaboration within a team of market researchers, consultants, and a manager who all work from their own homes. The system supported communication between a central office and remote workers and offered facilities for information storage, the configuration of a central database, and the preparation of client reports.

In Scotland, a system was developed between a large central hospital and a small clinic on a remote island (case 8 in Exhibit 1). The system is used to transmit high-quality X-ray images, together with other audio, visual, and text support, in real time in order to facilitate remote diagnosis by medical experts in the central hospital. Similarly, in Greece, a new system was developed to provide a full-time medical consultancy involving a major teaching hospital in a large urban area and some small clinical units based in remote rural areas (case 9 in Exhibit 1). The final case study was of a homework-based telephone enquires service in Portugal (case 10 in Exhibit 1) where a data network was used to support the management and supervision of home-based workers and to enable communication between co-workers in order to provide support and guidance and to help avoid problems of social isolation.

The creation of virtual teams in these case studies brought a range of benefits to the organizations and individuals concerned, including increased geographical flexibility for team members; more efficient and effective use of expert time; improved responsiveness; increased productivity; more satisfying working environments; and reduced costs. However, the process of creating and operating these teams was not problem-free, and to achieve their full potential, a number of barriers needed to be overcome.

In all of the case studies, there were a number of relatively straightforward technical problems, ranging from unreliable systems and incompatible networks to slow computers and poor response times at certain times of the day. However, the issues of trust and identity posed a more serious impediment to the effectiveness of these virtual teams. The issue of trust was most apparent where team members needed to share work-in-progress electronically. For example, the software developers in cases 2 and 3 were reluctant to share half-finished programs with others. Similarly, the consultants and market researchers (cases 6 and 7) were often unwilling to share half-written reports with colleagues.

Even when team members are prepared to share information and knowledge with each other, the time and effort required to manage communication can be a serious problem. This is mainly associated with the issue of identity: knowing whom one is in communication with, and precisely how that communication should be managed. This problem is often seen in medical settings, such as cases 8 and 9, where a number of different professions need to communicate across professional boundaries. For example, Kindberg, Bryan-Kinns, and Makwana (1999) describe the technical and professional trade-offs that clinicians need to make when they are unable to ascertain exactly who their reports will be seen by and in what context they will be read.

In the sections that follow, we will discuss some of the challenges faced in managing virtual teams under two broad headings. The first concerns problems that are in some way related to the technology that is used. These are not technological problems per se, but the human and managerial problems that the perceived deficiencies in the technology cause. The second are essentially organizational problems, where the issue is more concerned with people, and how they are managed in virtual environments.

## 5   Technological Problems

Establishing trust in online transactions is a well-recognized problem in e-commerce; it is also a significant challenge for virtual teams. Online communication lacks the richness of face-to-face interaction. Relying solely on online communication tends to inhibit participation and the creation of trust and the sense of mutual responsibility that characterizes teamwork. Jarvenpaa and Leidner (1999) have highlighted the importance of the first "online impression," which can set the tone for much of the later discussion. Jarvenpaa, Shaw, and Staples (2004) argue that trust has both a direct and a mediating effect on team effectiveness. Given this problem, most virtual teams will still tend to favor holding occasional face-to-face meetings. For example, a study by Kimble and Hildreth (2005) that dealt with a virtual team spread across three continents found that although the team had access to some of the most sophisticated and up-to-date communications technology available, the "boost" provided by face-to-face meetings was needed to sustain it through extended periods of online communication.

Studies have identified two principal modes of interaction in virtual teams (Sivunen & Valo, 2006). The first is sometimes termed "hot" virtual working and is also known variously as "closely coupled," "tightly coupled," or "online" working. This is interaction in the sense that we would normally think of it - that is, synchronous, fluid, and requiring the active participation of the other members of the team. The second is "cold" virtual working, which is sometimes termed "loosely coupled" or "off-line" working. This form of interaction is work in the sense that it is part of some collective activity directed toward a shared goal or common purpose, but it is work that is performed individually. In general, it does not require the active presence of the other members of the group and can be performed alone and asynchronously.

During the life of a team, the interaction between its members moves repeatedly between these modes of communication. For example, Ribeiro, Kimble, and Cairns (2010) noted how people in the groups they studied would sometimes use the techniques of cold distributed working even when they shared the same physical office space. However, while their use of technology was "fluid and almost transparent" (Ribeiro et al., 2010, p. 27), their decision about when to use it was not. The members of the groups were happy to use technology to maintain day-today contact, yet when it became too difficult or involved delicate decisions, face-to-face meetings were still the preferred method of communication. When face-to-face meetings are feasible, this need not be a problem, but in situations where they are either too costly or would be inefficient, as in the example of the software support teams in case 1, other solutions need to be found.

The technological solution to the problem of not being able to meet somebody face to face relies on the creation of what is called social presence. The classic definition of social presence is the degree of awareness of other people in an interaction and the subsequent recognition of interpersonal relationships (Short, Williams, & Christie, 1976). Now social presence, or co-presence, is more commonly taken to mean the degree to which one can form a sense that one is interacting with another individual. Early attempts to deal with this problem relied on trying to provide "contextual information" about the other person, usually in the form of textual descriptions; later attempts relied on avatars (computer-generated animated images) to simulate co-presence. Now this problem is most often dealt with by some form of teleconferencing. In case 1, where establishing a close working relationship quickly was critical, great efforts were made to create a sense of co-presence. However, in other cases, such as case 5, this was of less importance, and document management software that supported the simultaneous editing of documents was simply augmented by standard tele-conferencing facilities.

## 6   Organizational Problems

As we have seen, without effective communication effective teamwork becomes difficult. While some aspects of this problem can be dealt with by technological means, others are more fundamentally rooted in the ways in which people work and are managed. Identity, for example, plays a critical role in communication, where knowing the identity of those with whom you communicate is central to creating a shared understanding. Yet in virtual teams, the status and identity of a person can be ambiguous, as many of the basic cues that exist in the physical world may be absent; in the virtual world, one can have as many electronic personas as one has the time and energy to create. At the individual level, this can lead to the problems of trust outlined earlier, but at the level of the team, it can lead to difficulties in establishing a sense of collective identity and can inhibit effective communication among team members.

Identity helps to establish shared meanings through providing a common perspective on, for example, where somebody's job fits within the wider organization. Without this common ground, it becomes difficult to share knowledge effectively. At a more fundamental level, unless shared meanings can be established, even "common sense" words and terms become open to different interpretations, as the same word may be used in different ways in different settings. Identity in this sense is defined by a group rather than by an individual. Within a group, a shared interest or a common domain of knowledge provides the "common sense" definitions for the words and terms used by the group. Sharing that group's identity implies not only a commitment, but also shared knowledge and shared competencies. Teams are mainly focused on achieving a task rather than building a sense of identity. Consequently, we need to look elsewhere for a different type of group structure if we are to solve the problems of building and maintaining group identity.

As Exhibit 2 illustrates, Communities of Practice are a different cut on the organization's structure that emphasizes the learning that people do rather than the functional unit they report to or the project they are working on (Wenger, 1998). Because membership in a Community of Practice is based on a shared interest, it can cross organizational boundaries and span structures and hierarchies. Communities of Practice are "not just places where local activities are organized, but where the meaning of belonging to broader organizations is negotiated and experienced" (Wenger, 1996, p. 25). Thus, for example, people who work in cross-functional teams might form a Community of Practice to keep in touch with their peers and keep up to date with what is happening in their particular field of specialist expertise.

Richard McDermott coined the phrase "the double-knit organization" (McDermott, 1999) to describe an organization that combines teams with Communities of Practice and by doing so overcomes some of the problems of maintaining a sense of identity online. He argues that cross-functional teams focus on outputs such as products, processes, or market segments, while Communities of Practice focus on working together to solve shared problems, to learn, and to build a body of knowledge. He sees Communities of Practice as a way to maintain a technical focus within a broader discipline, while cross-functional teams serve to unite those disciplines around a common product. Instead of sharing product or process-specific information via team leaders, Communities of Practice share knowledge and standardize practices across teams.

Exhibit 2. Comparison of Communities of Practice and Teams (Adapted From Wenger and Snyder, 2000)

|  | **What's the purpose?** | **Who belongs?** | **What holds it together?** | **How long does it last?** |
|---|---|---|---|---|
| **Community of Practice** | To develop members' capabilities and to build and exchange knowledge | Members select themselves | Passion, commitment, and identification with the group's expertise | As long as there is interest in maintaining the group |
| **Team** | To accomplish a specified task | Employees assigned by senior management | The project's milestones and goals | As long as the project |

Other research has also indicated that Communities of Practice may be one way to make some inroads into the complexities and challenges of virtual working. Pemberton-Billing, Cooper, Wootton, and North (2003) used the concept of a Community of Practice to highlight some of the root causes of the problems they found in their study of distributed design teams, such as the problems that a hierarchical client/supplier relationship caused when attempting to create a common sense of purpose for the team. Similarly, research by Hildreth (2000) has indicated that the willingness to go "the extra half mile" in a Community of Practice can help overcome many of the problems associated with issues of trust and identity in virtual working and enable relationships in virtual teams to develop quicker, go further, and provide a sound basis for subsequent hot and cold electronic collaboration.

# 7 Conclusions

The goal of this article was to discuss some of the challenges that organizations face when managing virtual teams. One of the main conclusions must be that there are no easy answers. To ensure that virtual teams work effectively, you need to address both the people issues and the technology, rather than look for the answer in one or the other. Davis refers to this as "the Tao of leadership in virtual teams" (Davis, 2004) and describes managing virtual teams as managing a paradox. He notes, "Numerous paradoxes exist in virtual teams. Attempts to focus on one side of the paradox yield only limited success" (Davis, 2004, p. 57). Some problems can be ameliorated by improving or updating technology, and others by finding new ways to organize workgroups. However, truly effective virtual teams will only be built upon understanding the limitations of virtual working on a human scale and finding ways, both technological and managerial, to overcome them.

While it is clear that, because virtual teams are composed of groups of people who must work together as a single cohesive entity, technology alone can never provide all of the solutions, the role of technology cannot be ignored. Thus, Breu and Hemingway (2004) describe how an attempt to create virtual teams in the public sector floundered because of an unreliable and inadequate technological infrastructure. Similar problems can be found in the private sector, where a visit to a client's site can mean that resources that are readily available in the headquarters are no longer accessible.

Although technology cannot be ignored, it is not always the case that particularly sophisticated technology is needed. For example, Henttonen and Blomqvist (2005) describe how, in a global virtual team where most of the members had never met each other, traditional means of communication such as telephone and e-mail were more popular than Web based collaborative tools and groupware. It seems strange that although the concept of a virtual team only became reality thanks to advances in information technology, the technology itself often seems to play but a small part.

A different solution to the problems of virtual working is to blend virtual teams and Communities of Practice in a "double-knit" organization. Even here, we find a paradox. Just as McDermott concedes that teams can become the new functional silos because a team's focus on fulfilling a task can lead to isolation and team myopia (McDermott, 1999), so others have found that his solution to the problem, Communities of Practice, can have the same shortcomings. Hislop (2003), for example, found that certain types of knowledge, although valuable to others, never moved beyond the boundaries of the community. Similarly, while the self-managing characteristics of teams can be a bonus, the same characteristic in a Community of Practice can be a problem. Gongla and Rizzuto (2004), for example, found numerous examples of Communities of Practice that simply "disappeared" when the interests of the community and the interests of the organization diverged. While some of the theoretical concepts from Communities of Practice might help to inform and give insights into the way that virtual teams operate, Communities of Practice alone are not the whole solution.

The desire to communicate and work with others in groups is part of human nature, but the rapid development of communications technology, in all of its forms, has added a new dimension to this basic desire. Within a single generation, we have moved from fixed location, one-to-one communication by telex and telephone to a whole range of different possible modes of communication, all of which have been opened up by the sudden and rapid expansion of digital networks. Perhaps, to paraphrase Mark Twain, Cairncross's reports of "the death of distance" (Cairncross, 1997) have been greatly exaggerated, but there is no doubt that, technically at least, working at a distance is no longer the challenge that it once was. The challenge to us as human beings is in how we adapt and respond to these new opportunities.

# 8 References

Breu, K., & Hemingway, C. J. (2004). Making organisations virtual: The hidden cost of distributed teams. Journal of Information Technology, 19(3), 191–202.

Cairncross, F. (1997). The death of distance: How the communications revolution will change our lives. Boston: Harvard Business School Press.

Davis, D. D. (2004). The Tao of leadership in virtual teams. Organizational Dynamics, 33(1), 47–62.

Gongla, P., & Rizzuto, C. R. (2004). Where did that community go? Communities of practice that "disappear." In P. Hildreth & C. Kimble (Eds.), Knowledge networks: Innovation through communities of practice (pp. 295–307). London: Idea Group Publishing.

Gyllenhammar, P. G. (1977). People at work. Reading, MA: Addison-Wesley.

Henttonen, K., & Blomqvist, K. (2005). Managing distance in a global virtual team: The evolution of trust through technology-mediated relational communication. Strategic Change, 14(2), 107–119.

Hildreth, P. (2000). Going the extra half-mile: International communities of practice and the role of shared artifacts (Unpublished doctoral dissertation). University of York, York, U.K.

Hislop, D. (2003). The complex relations between communities of practice and the implementation of technological innovations. International Journal of Innovation Management, 7(2), 163–188.

Jarvenpaa, S. L., & Leidner, D. E. (1999). Communication and trust in global virtual teams. Organization Science, 10, 791–815.

Jarvenpaa, S. L., Shaw, T. R., & Staples, D. S. (2004). Toward contextualized theories of trust: The role of trust in global virtual teams. Information Systems Research, 15, 250–264.

Kimble, C., Grenier, C., & Goglio-Primard, K. (2010). Innovation and knowledge sharing across professional boundaries: Political interplay between boundary objects and brokers. International Journal of Information Management, 30, 437–444.

Kimble, C., & Hildreth, P. (2005). Dualities, distributed communities of practice and knowledge management. Journal of Knowledge Management, 9(4), 102–113.

Kimble, C., Li, F., & Barlow, A. (2000). Effective virtual teams through communities of practice (Department of Management Science Research Paper Series No. 00/9). Strathclyde, UK: University of Strathclyde.

Kindberg, T., Bryan-Kinns, N., & Makwana, R. (1999). Supporting the shared care of diabetic patients. Paper presented at the ACM SIGGROUP Conference on Supporting Groupwork.

McDermott, R. (1999). Learning across teams: The role of communities of practice in team organizations. Knowledge Management Review, 8, 32–36.

Pemberton-Billing, J., Cooper, R., Wootton, A. B., & North, A. N. W. (2003). Distributed design teams as communities of practice. Paper presented at the 5th European Academy of Design Conference, Barcelona.

Ribeiro, R., Kimble, C., & Cairns, P. (2010). Quantum phenomena in communities of practice. International Journal of Information Management, 30(1), 21–27.

Short, J., Williams, E., & Christie, B. (1976). The social psychology of telecommunications. London: Wiley.

Sivunen, A., & Valo, M. (2006). Team leaders' technology choice in virtual teams. IEEE Transactions on Professional Communication, 49(1), 57–68.

Wenger, E. (1996). Communities of practice: The social fabric of a learning organization. Healthcare Forum Journal, 39(4), 20–24.

Wenger, E. (1998). Communities of practice: Learning as a social system. Systems Thinker, 9(5), 1–5.

Wenger, E., & Snyder, W. (2000). Communities of practice: The organizational frontier. Harvard Business Review, 78(1), 139–145.